BMC
Evolutionary Biology

# A Boolean gene regulatory model of heterosis and speciation

Peter Martin Ferdinand Emmrich[1,2*], Hannah Elizabeth Roberts[1,3] and Vera Pancaldi[1,4]

## Abstract

**Background:** Modelling genetic phenomena affecting biological traits is important for the development of agriculture as it allows breeders to predict the potential of breeding for certain traits. One such phenomenon is heterosis or hybrid vigor: crossing individuals from genetically distinct populations often results in improvements in quantitative traits, such as growth rate, biomass production and stress resistance. Heterosis has become a very useful tool in global agriculture, but its genetic basis remains controversial and its effects hard to predict. We have taken a computational approach to studying heterosis, developing a simulation of evolution, independent reassortment of alleles and hybridization of Gene Regulatory Networks (GRNs) in a Boolean framework. These artificial regulatory networks exhibit topological properties that reflect those observed in biology, and fitness is measured as the ability of a network to respond to external inputs in a pre-defined way.

**Results:** Our model reproduced common experimental observations on heterosis using only biologically justified parameters, such as mutation rates. Hybrid vigor was observed and its extent was seen to increase as parental populations diverged, up until a point of sudden collapse of hybrid fitness. Thus, the model also describes a process akin to speciation due to genetic incompatibility of the separated populations. We also reproduce, for the first time in a model, the fact that hybrid vigor cannot easily be fixed by within a breeding line, currently an important limitation of the use of hybrid crops. The simulation allowed us to study the effects of three standard models for the genetic basis of heterosis: dominance, over-dominance, and epistasis.

**Conclusion:** This study describes the most detailed simulation of heterosis using gene regulatory networks to date and reproduces several phenomena associated with heterosis for the first time in a model. The level of detail in our model allows us to suggest possible warning signs of the impending collapse of hybrid vigor in breeding. In addition, the simulation provides a framework that can be extended to study other aspects of heterosis and alternative evolutionary scenarios.

## Background

Heterosis, or hybrid vigor, is the phenomenon by which the offspring of crosses of individuals from compatible but genetically distinct populations of the same or different species show enhanced characteristics valuable to breeders [1]. Increases in fertility, biomass production, grain number and total yield have been described [2].

* Correspondence: peter.emmrich@jic.ac.uk
[1]Department of Plant Sciences, University of Cambridge, CB2 3EA Cambridge, UK
[2]Current address: John Innes Centre, Norwich Research Park, Norwich NR4 7UH, UK
Full list of author information is available at the end of the article

This effect has been known for a long time among breeders, and scientific descriptions date back to Charles Darwin [3]. One important aspect of heterosis is the difficulty of fixing the favorable properties of the hybrids into a true breeding line, making repeated crosses from inbred lines the most practicable method of harnessing the benefits of heterosis. Hence the study of heterosis is of economic as well as biological significance. The phenomenon has long been utilized in the breeding of crops such as maize [4] and is also known to be important in natural evolution, through inbreeding depression. Recent progress in understanding the underlying genetic and metabolic factors have provided breeders with methods to predict levels of heterosis [5]. With the advent of cheaper genome-wide profiling technologies, both at the RNA and proteomic level, a more quantitative picture of heterosis is emerging, with multiple studies reporting regulation of gene expression and protein levels that differs between parents and hybrids [6,7].

Nevertheless, heterosis remains difficult to predict in the absence of complete metabolic or transcriptomic profiles: not all divergent crosses produce heterosis and there is little correlation between heterotic effects observed for different biological traits. For some traits genetic distance is not a good a predictor of heterosis [8].

The genetic mechanisms underlying heterosis are still controversial. In the first half of the twentieth century several hypotheses were put forward: According to the dominance model [9], heterozygosity is advantageous because hybrids receive a beneficial dominant alleles (which mask deleterious recessive alleles that are homozygous in one of the two parents) for a greater number of genes than either of their parents. In contrast, the model of over-dominance suggests that the presence of heterozygosity at certain loci is advantageous in itself [10]. According to the third model, pseudo-over-dominance, heterosis occurs if the hybrid inherits reciprocal linked pairs of beneficial dominant and harmful recessive alleles from its parents [11]. In this case, the advantage is conferred by the presence of both dominant alleles, but due to linkage, the dominant alleles cannot be readily recombined. In breeding, cases of pseudo-over-dominance will generate a very similar segregation pattern to over-dominance [12]. In addition, there may be epistatic effects [13], where the effect of an allele at one locus depends on the allele or alleles present at other loci, e.g. an allele improves fitness only if a certain other allele is present at another locus.

These models are non-exclusive and all four mechanisms may contribute to heterosis [14]. Their relative importance remains the subject of intense discussion and may differ between organisms [1,5,15-18]. Whilst evidence suggests that all the above genetic models may play a role in heterosis, they may not provide a full explanation of the phenomenon. Recently, some researchers have suggested putting them aside entirely, in favor of analyzing the quantitative traits that underlie heterosis using systems biology approaches, focusing on regulatory interactions and multigenic effects [1,5,19-22].

Accordingly, Andorf and colleagues proposed a network model of heterosis which they initially presented in a simulation study and then successfully validated using experimental data [5,19,20]. Their main hypothesis is that hybrids perform better than parents because there are more regulatory interactions in the hybrid networks representing the individuals than in those of the parent and this additional complexity increases adaptability in changing environments. They define the networks based on partial correlations between experimentally measured gene expression and metabolite levels and they demonstrate that these networks are more densely interconnected in hybrids. Such findings relate the topology of the network (a systems property) to heterosis.

Simpler models considering the role of macromolecular complexes and dominant negative interactions have also been proposed as explanations for heterosis [23].

Unfortunately, large scale experimental studies over evolutionarily relevant timescales and with detailed network reconstruction for each individual are not feasible [24]. We have therefore produced a realistic simulation of such a study in which we aimed to reproduce the following well-studied experimental observations associated with heterosis [25-27]:

1. The offspring of crosses of individuals from separate populations perform better than inbred offspring in several quantitative traits, i.e. they show heterosis.
2. The heterotic effect increases with increasing genetic divergence of the parents.
3. This increase occurs up to a limit after which divergence is too great and parents are not able to produce viable offspring.
4. The heterotic effect cannot be fixed in a genetic lineage by repeated intercrossing of heterotic hybrids.

Inspired by literature on Gene Regulatory Networks (GRNs) [28-32] and evolutionary systems simulations [33], we simulated evolution and hybridization of GRNs using Boolean networks. Boolean network modeling was first applied to the study of GRNs in the 1960s [34] and has since been used extensively for the simulation of networks in molecular biology [35-41]. Kauffman's approach considered each cell to be represented by a random network and associated the attractors of the Boolean network to the different cell-types, thus explaining how a single genotype (the network) could lead to multiple phenotypes (the attractors would be seen as different cell types in a multi-cellular individual). Recently, a three-gene Boolean regulatory network was used to simulate a genotype-phenotype relationship in a study of multi-functionality in biological systems, demonstrating the potential of the framework [42].

In our approach we similarly modeled the genome and regulatory interactions within an individual as a Boolean network. The dynamics of this network, that is the changes of the network node states, represents the regulatory response of the network to environmental inputs, while the final stable state reached by the network – its attractor – represents the phenotype. The Boolean model assumes that a gene is either ON or OFF and that the quantitative changes seen in real world GRNs can be produced by the state changes generated by the regulatory interactions. We believe this is a reasonable simplifying assumption and the Boolean logic used to simulate the dynamics of our networks can reproduce important characteristics of biological regulatory networks and allow the simulation of sufficient individuals to study evolutionary processes within reasonable computational time. We were able to model the evolution of reasonable size networks (up to 50 nodes) in each individual of a population and allowed the population to evolve over thousands of generations. The quantitative fitness score that was used to select individuals that contributed to the next generation depended on their ability to respond to several sets of simulated environmental factors.

Our simulation was successful in reproducing the four experimental observations outlined above. It was able to generate hybrid vigor for offspring of crosses between populations that had been allowed to diverge for several generations, and it also displayed a dramatic fall in fitness as the genetic distance between parents increased to the point of preventing the formation of viable offspring. We thus recapitulate the results of the model of Andorf et al., which showed that there is a maximum level of heterosis that cannot be exceeded by increasing genetic distance between the parents [20] and the related experimental observations that genetic incompatibility can cause reduced fitness of hybrids [25,27]. Our model goes beyond this earlier work by demonstrating that hybrid vigor is quickly lost when members of the same heterotic hybrid population are crossed, a well-known biological phenomenon of great consequence for breeders. In addition, the architecture of our networks and our evolutionary algorithm are designed to more closely resemble the evolution of gene regulatory networks found in nature. Our model thus allows us to investigate the different mechanisms underlying multigenic heterosis, including epistatic interactions. In summary, by simulating the evolution of gene regulatory networks in response to environmental inputs, we have constructed a tool for the *in silico* investigation of hybrid vigor and its determining mechanisms.

## Results

Our model simulates evolution and hybrid formation for diploid organisms, using gene regulatory networks and their response to environmental inputs.

On the individual level, we simulated the genetic and regulatory characteristics of each diploid individual as a Boolean network. Each phenotype is defined as the regulatory steady state that is reached by a network in response to inputs from outside the network (environmental factors). We calculate fitness as the similarity of the network phenotype (dynamically stable state) to the ideal phenotype under certain environmental conditions.

In order to prevent the predominance of networks which simply reach the perfect output for a single specific condition, with no advantage in maintaining the regulatory dynamics, we subject each network to three sets of environmental factors (three environments), each requiring a specific response. The network fitness is a combined measure of the appropriateness of the responses to each set of environmental factors.

At the population level, we simulate the evolution of a set of individuals (each represented by a network) through cycles of mutation, reassortment and selection in a set of specific environments. First, we allow several hundred generations for the population to adapt to the environments. Then we duplicate all networks and then let them evolve separately, forming two, initially identical populations that diverge, since they are not allowed to interbreed (Figure 1). For the details of the simulation we refer the reader to the Methods section and Additional file 1.

### Population fitness increases stochastically

At the beginning of the simulation the population fitness is close to zero, because the network dynamics of naïve individuals are entirely random. Whilst we generate networks to display certain quasi-biological characteristics in their topology - such as a scale-free-like degree distribution [43] and clustering [44] - the state of each node is just as likely to be the correct response to a given set of environmental factors as it is to be the opposite. Since fitness is averaged over nodes responding correctly and nodes responding incorrectly, they tend to cancel out in a network that has not undergone adaptation, leading to the observed fitness measure around zero. However, after mutations through evolutionary time, individuals are produced with a response closer to the optimal one. These individuals are selected to produce offspring, which increases the average fitness of the population over time.

As can be seen in Figure 2, the population fitness increases through evolutionary time, with a characteristic pattern of punctuated equilibria. This feature of the graph represents the fact that improvements in fitness are brought about by rare but important mutations that produce sudden increases, alternated by periods of more or less constant fitness where only neutral or deleterious mutations occur. The latter are quickly removed from the population by selection, causing them to have little effect on the fitness of the overall population. In the very beginning, all individuals in the population are identical, but during the adaptation period, the diversity increases. This causes the population fitness to become more variable over time in some runs of the simulation.

### Hybrids display higher fitness than intra-population crosses

After the networks have reached a fitness plateau with only infrequent increases in fitness (typically after 1000 generations), we duplicate the population and continue the evolution of both copies separately. Although the environments of the two populations remain the same, these two sets of individuals start diverging, as there is no genetic exchange between them. The evolutionary algorithm continues on both populations in the same way as before the separation, so fitness increases may still occur in either of the separated populations. Every few generations, we compared the fitness of offspring of crosses between individuals of the same population and separate populations, i.e. hybrids. The fitness of forty individuals of each parent population and eighty hybrid individuals was measured and averaged to generate the fitness values shown in Figure 3.

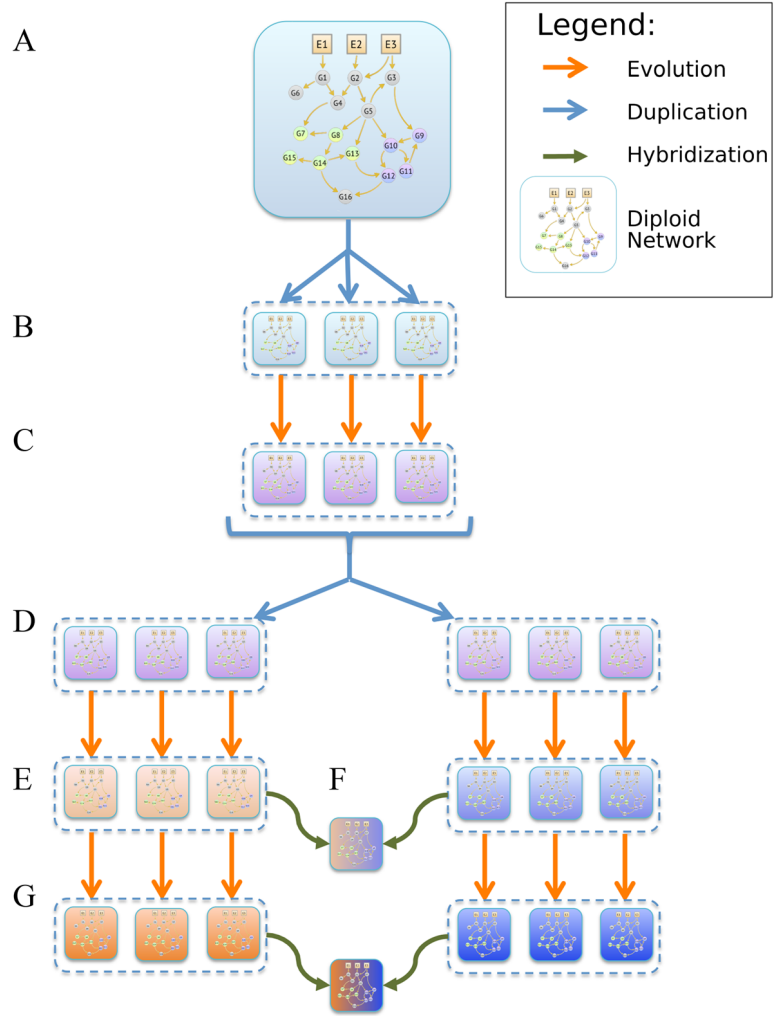

**Figure 1 Scheme of the evolutionary algorithm and the hybrid formation.** The initial network **(A)** is duplicated to form the initial population **(B)**, which evolves over hundreds of generations by mutation, reassortment and selection, to form an adapted population **(C)**. This population is duplicated, generating two initially identical populations **(D)**. Both populations carry on evolving, but do not interbreed, causing them to diverge (indicated by the orange and blue colors) **(E)**. At regular intervals, individuals are hybridized between the populations and the fitness of the hybrids is measured **(F)**. The hybrid individuals are then discarded and are not used for further breeding. Over time, the two parent populations diverge further (indicated by deeper shading) **(G)**.

As seen in real-world heterosis [8], the hybrid crosses displayed higher fitness than the intra-population ones in most, but not all cases and the amount of heterosis tended to increase through evolutionary time, presumably because the parent populations continued diverging from each other (see Figure 3A).

## The fitness of hybrids collapses after prolonged separation of parent populations

After a variable number of generations of separation between the parent populations, the fitness of the hybrids starts decreasing (Figure 3A). This precipitates a rapid collapse in hybrid fitness over the next few (usually less than 50) generations. After this, the hybrid fitness remains at a stable level far below the fitness of the parent populations. The time at which this collapse occurs varied greatly, between 20 and several hundreds of generations after the separation of the parent populations.

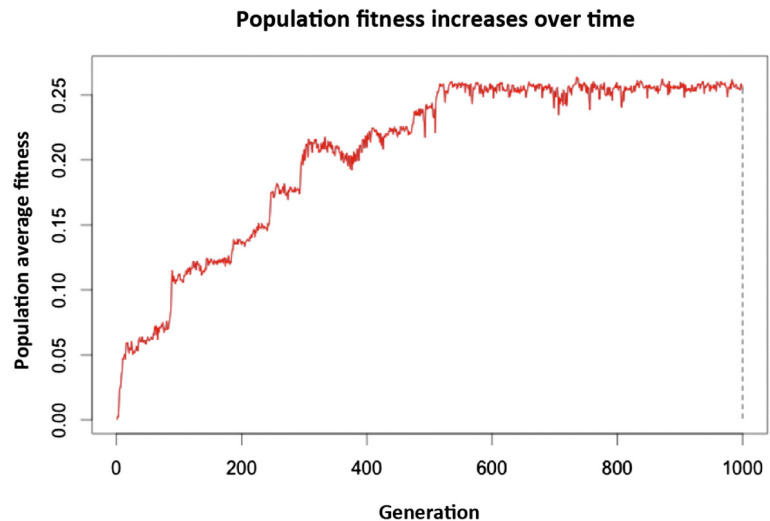

**Figure 2 Example of a typical adaptation graph, showing three stages.** Population fitness increased rapidly over the first five generations and more slowly over the next 500. This is a stochastic increase that sometimes occurs in sharp jumps in fitness, reminiscent of punctuated equilibria. After this, the population fitness varied around an equilibrium value for 500 generations.

### Heterosis does not persist when heterotic hybrids are crossed

We tested the fitness of crosses of between hybrids, to establish whether second and subsequent generation hybrids could preserve the increased average fitness. As can be seen in Figure 3B, the average fitness was not preserved in the subsequent generations, and instead quickly decayed back to the levels of the original parent populations. This is consistent with what is observed in nature, where heterotic phenotypes cannot be easily fixed into true-breeding lines [45].

### Categorizing the mechanisms underlying heterosis

Dominance and over/under-dominance (which we call local heterozygosity effects) involve mechanisms that operate at individual loci, whereas epistasis involves non-additive interaction between loci. In the dominance model, best-parent heterosis (in which the hybrid performs better than both parents) requires the additive effect of dominance at several loci (for an example see the Additional file 1: Text S1, Table S1). Over-dominance can produce best-parent heterosis even if it is only active at one locus. Our algorithm measures the strength of local heterozygosity effects at each heterozygous locus and the strength of epistatic effects for each pair of loci. The simulation allowed us to test the role of these different theoretical mechanisms of heterosis in contributing to fitness.

The results of an example run of the simulation are shown in Figure 4. Figure 4A shows the fitness performance of the two parent populations and the fitness of hybrids formed between them, while Figure 4B and Figure 4C respectively show the influence of local heterozygosity mechanisms and epistasis mechanisms on the fitness of the hybrids. Local heterozygosity effects are calculated by adding up the fitness effects of individual loci (disregarding epistasis). Epistasis effects are calculated by adding up the effect of each combination of two alleles on separate loci (disregarding epistasis involving more than two loci). This is only to allow a comparison between the strengths of different local heterozygosity effects or different epistasis effects, and does not allow an absolute estimate of the influence of these mechanisms on fitness. Therefore, we speak of 'relative fitness effects' in the following.

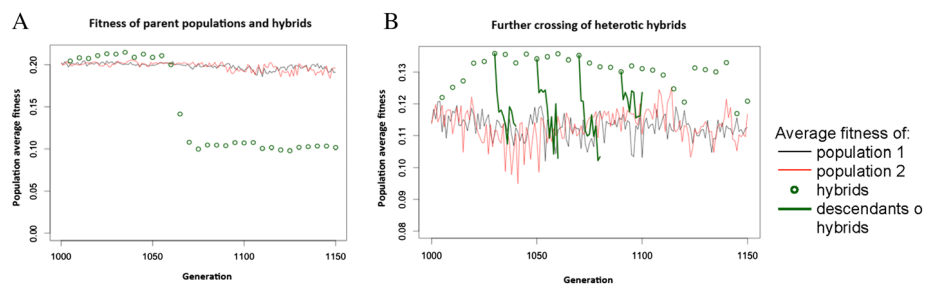

**Figure 3 Population fitness over time. A)** Hybrids perform better on average than the separated parent populations for some generations. After a variable number of generations, the fitness of newly formed hybrids collapses and does not recover. **B)** If hybrid individuals that exhibit heterosis are crossed to each other, the fitness advantage does not persist. Instead, the fitness of the offspring of hybrids collapses back to the same level as the parent populations. All fitness values are population averages of 40 (parent populations) or 80 (hybrids) individuals.

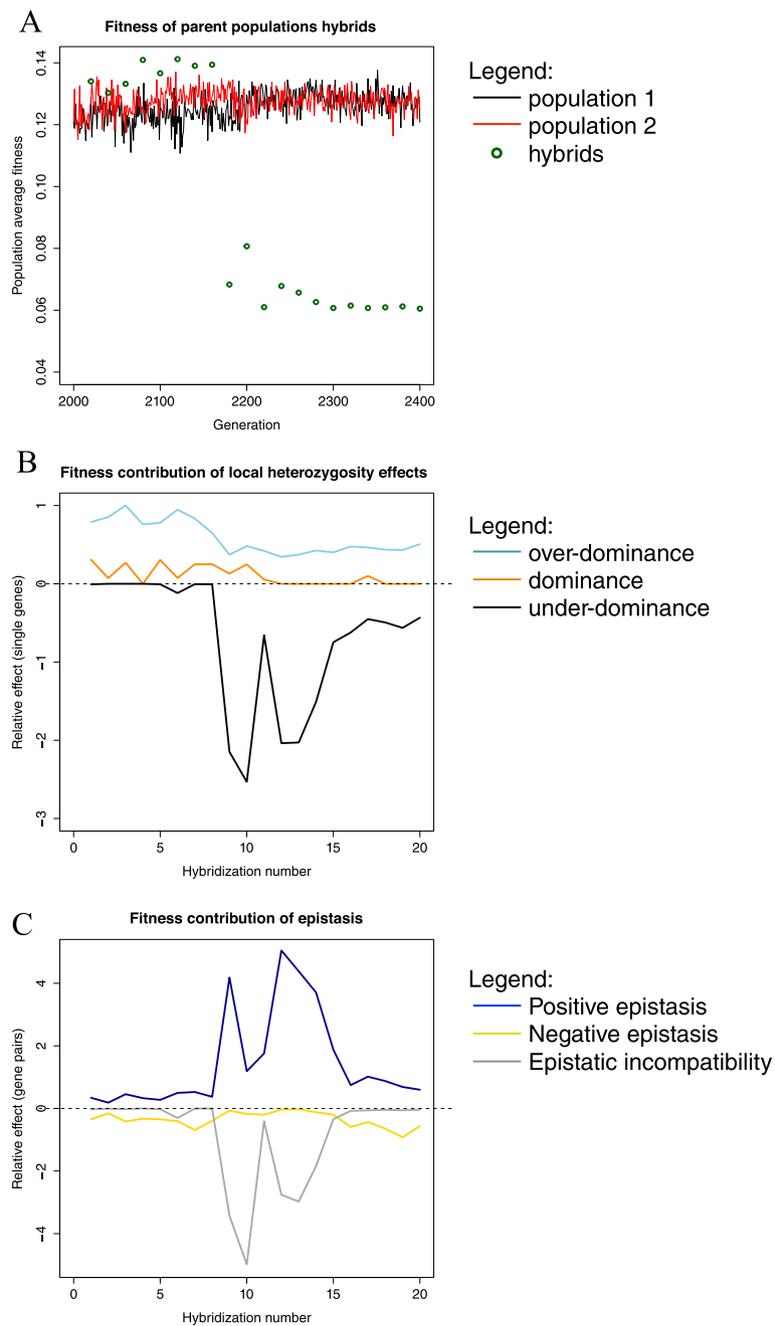

**Figure 4 Different mechanisms and their impact over time. A)** Performance of parent populations and hybrids in an example run. The parent populations evolved separately from generation 2000 onwards. Hybrids showed heterosis until generation 2150 followed by the collapse of fitness in the hybrids. Hybridization was performed every 20 generations; red and black: average fitness of parent populations, green circles: average fitness of hybrid population **B)** Effect of local heterozygosity mechanisms relative to each other; light blue: over-dominance, orange: dominance, black: under-dominance. **C)** Effect of epistatic mechanisms relative to each other; dark blue: positive epistasis, yellow: negative epistasis, gray: epistatic incompatibility.

### Relative contributions of local heterozygosity mechanisms to heterosis

Figure 5 shows data from 11 runs of the simulation (880 networks per time-point) for hybrid populations exhibiting heterosis, undergoing fitness collapse, and reaching a low fitness level after the collapse. The randomness of the process causes the collapse to occur very late in some cases (average 5.6 hybridizations, standard deviation 3.4). Hybridisations were performed at 20-generation intervals.

We calculated the relative strength of the effects of dominance, over-dominance and under-dominance (Figure 5A); see Methods for details on the calculation of each of these effects.

The effects of dominance are minor up until the point of collapse in hybrid fitness. During and after the collapse they become stronger, but remain very variable. In most networks, dominance does not contribute to the fitness at all. Over-dominance has a very strong effect right from the start in almost all networks. Under-dominance is absent at the start, but rapidly increases in strength (i.e. its effect becoming more negative) at the point of collapse. Curiously, it then weakens again, but remains stronger than dominance and over-dominance combined.

### Relative contributions of epistatic mechanisms

When we calculate epistasis, we consider the interaction of the paternal allele of one locus with the maternal allele of another. The strength of the effects of epistasis cannot be directly compared to the values of dominance, over-dominance and under-dominance because epistatic effects can involve alleles on more than two loci. Calculating the effects of epistasis between alleles on three or more loci would be prohibitively computationally challenging. In a case where there is a positive epistatic interaction between alleles on three loci (e.g. three nodes in a regulatory chain), where two are from parent 1 and one is from parent 2, losing just one of them would cause the loss of the function provided by their interaction. If only alleles on pairs of loci are considered for epistasis, this one instance of epistasis would be counted twice (once for each parent 1 allele paired with the allele of parent 2). Therefore, summing the effects of epistasis of loci pairs is likely to overestimate the true strength of epistasis and its contribution to heterosis. However, the effects of different kinds of epistasis can be compared to each other (Figure 5B).

Positive epistasis (e.g. the two genes considered are part of the same pathway) is weak before the collapse and remains at an intermediate level after. Negative epistasis (e.g. genes are part of parallel pathways) is relatively weak throughout the simulation. Epistatic incompatibility (meaning that the presence of both alleles reduces fitness) starts off near zero, but becomes very strong during the collapse and then dissipates again to an intermediate level, following the same pattern as under-dominance.

In our simulation, positive epistasis is almost absent before the collapse and dominance is weak and over-dominance appears to be the main contributor to heterosis. The collapse of hybrid fitness is caused by both under-dominance and epistatic incompatibility.

### Early signs of hybrid fitness collapse

The collapse in hybrid fitness occurs very suddenly and there is no apparent slow build-up of deleterious mutations. Instead, the patterns of under-dominance and epistatic incompatibility suggest that individual mutations cause the fitness collapse by sweeping through one of the parent populations. The mutation must be beneficial in the population in which it arose or else it would not be selected. In the hybrids, however, it causes strongly reduced fitness. The transition from high (heterotic) fitness to low fitness is binary for each individual network. Thus, even before the average fitness of the hybrids drops catastrophically, there are individual hybrid networks with very low fitness. These alleles would not normally be removed by selection, as their effects in the background of the parent populations are neutral or positive. The appearance of hybrid individuals with very low fitness can indicate imminent incompatibility, if the responsible alleles sweep through the parent population.

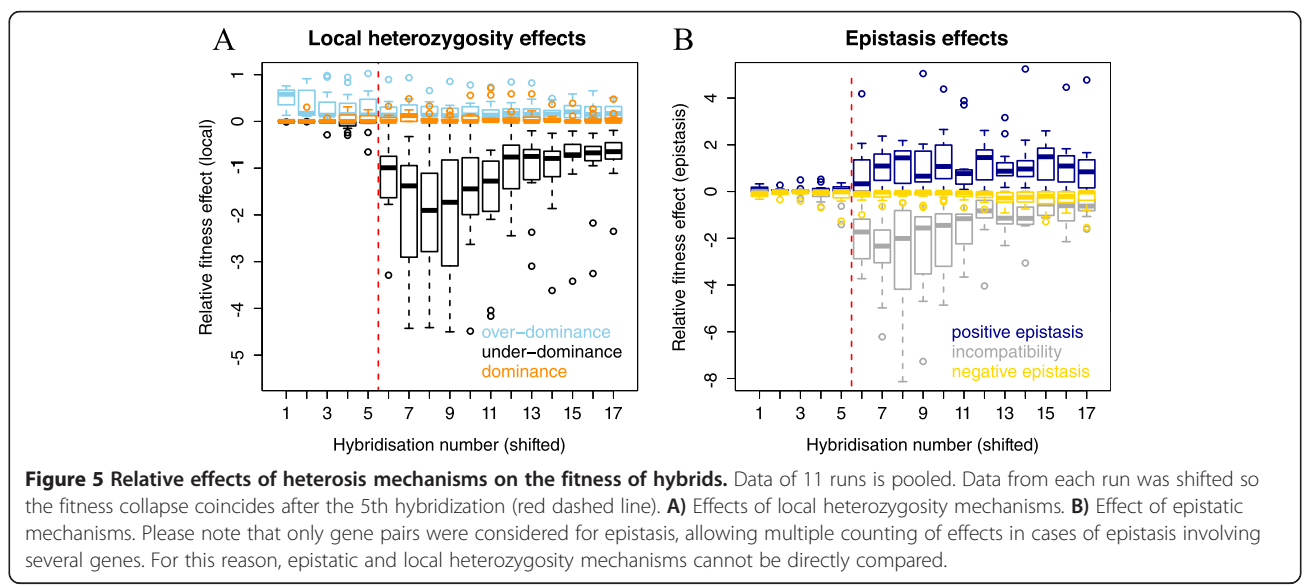

**Figure 5 Relative effects of heterosis mechanisms on the fitness of hybrids.** Data of 11 runs is pooled. Data from each run was shifted so the fitness collapse coincides after the 5th hybridization (red dashed line). **A)** Effects of local heterozygosity mechanisms. **B)** Effect of epistatic mechanisms. Please note that only gene pairs were considered for epistasis, allowing multiple counting of effects in cases of epistasis involving several genes. For this reason, epistatic and local heterozygosity mechanisms cannot be directly compared.

## Discussion

We have presented a model that can reproduce many biological observations associated with hybrid vigor [1,8]. This includes the observation that vigor increases as the parents diverge genetically up to a point where the parent populations become incompatible [46] and hybrids between them are at a severe disadvantage, as well as demonstrating the difficulty of fixing the hybrid trait in a true-breeding line [45].

Our model considers diploid organisms and it implements independent reassortment of alleles. We take a similar approach to the network model of Andorf et al. [5,19,20] in that our phenotype (and hence the levels of heterosis) are directly related to the networks' topologies. However, we have a more sophisticated definition of the regulatory interactions between the genes, which makes our model more biologically realistic. The main focus of Andorf's work evolves around the assumption that a significant component of heterosis is determined by structural properties of the network. We do not explicitly use this and actually trying to test this hypothesis we observed that for small networks, like the ones we can consider (30–50 nodes), the definition of network statistical properties is not particularly meaningful. Furthermore, we have used simulations of entire populations of individuals evolving in time to explore the emergence of heterosis at the population level, allowing us to investigate the importance of the different genetic mechanisms that have been suggested to explain the phenomenon.

Moreover, we have provided a novel definition of fitness, which is based on a combination of the individual's genotype and environment. The approach represents a compromise between a very detailed simulation of the genotype or of the phenotype. For example, [47] model the emergence of functional gene clustering explicitly considering the physical distribution of genes along chromosomes, accounting for linkage and other factors as well as considering the corresponding protein interaction network. However they employ a definition of fitness which is very different from ours and mostly focuses on the individuals' network characteristics rather than their interaction with the environment (based on optimizing the number of cells or the number of cell types). On the other hand, other interesting models have used a more sophisticated definition of phenotype, sacrificing accuracy in the representation of the genotype [48], but these were not aimed at studying heterosis. Our work focuses on a realistic representation of the gene regulatory dynamics, which gives rise to a natural definition of the phenotype, importantly, related to the environment.

Although we don't explore this here, this framework could be used to study the impact of different types of environment in shaping the phenotype. For example, the environments in which the individuals are selected could be changed periodically and the time scale of these changes is likely to affect the evolution of the populations and the extent of the hybrid vigor observed. Further investigation should lead to the identification of features of the environment and of the regulatory network of the parents that determine the extent and robustness of the hybrid vigor. In addition, further work will aim to identify the signs of the impending collapse in hybrid vigor.

### Continued separate evolution produces a speciation-like event

We hypothesized that the two parent populations would diverge over time, as each accumulates unique mutations. While this genetic difference between the populations gives rise to heterosis [49]. Yet, once two parent individuals have evolved to become so different that combining their regulatory networks causes problems, the fitness of hybrids will be reduced. We thus expected to see heterosis to decrease again after the parent populations had evolved separately for some time. In the simulation, we observed this decrease as a drastic collapse of hybrid fitness. It is worth noting that this is not an intended function of the simulation, but rather an emergent behavior of the evolutionary algorithm. The number of generations after the population split at which this collapse occurred varied widely, between ten and several hundred generations. In biological terms, this would describe an event akin to speciation [50]. Hybrids that form between the two populations after this event would be at a severe disadvantage (Figure 3A). Although our model does not have a spatial component, the two populations are prevented from interbreeding, as they would be through spatial separation. After the collapse, the separation of the two populations would be reinforced by their genetic incompatibility, even if they were to mix again. It should be noted that the two populations, while being kept separate, were subject to the same environments and were under exactly the same selective pressure. This observation closely resembles the findings of the ecological study on the fitness of intra- and inter-population hybrids performed by Drury and Wade [51]. As in our simulation, individual genes that had emerged in the separated populations of red flour beetles had weak advantageous effects in the background of their own population, but caused a drastic fitness reduction in hybrids.

It is interesting to note the sharpness of the hybrid fitness collapse. In most cases, the average hybrid fitness dropped to a steady level far below the fitness of the parents in a space of less than ten generations. This suggests that a very small number of mutations cause the incompatibility between the parent populations. These mutations may be advantageous in one of the parent populations and sweep through it rapidly. In conjunction with genes from the other parent population, however, they disrupt the regulatory network, leading to low fitness. It is possible that some of the mutated alleles that cause heterosis are also involved in causing the fitness collapse, but our simulation does not presuppose this.

### Over-dominance is an important mechanism of heterosis in our model

In our simulation, over-dominance is the mechanism with the greatest contribution to network fitness during the generations in which heterosis is observed. We are aware that this result could be influenced by the assumptions of our model, such as the binary nature of Boolean networks and our definition of fitness and environments. Whether and to what extent overdominance contributes to heterosis observed in biology is a highly controversial topic [14]. Experimental evidence in support of the overdominance model has been found in rice [52] reviewed in [15], sorghum [53], tomato [54] and oil palm [55], among others. In the model, under-dominance and epistatic incompatibility both contribute to the observed collapse in hybrid fitness. Positive epistasis and epistatic incompatibility almost always appeared together, with their increasing and decreasing fitness effects directly opposite to each other and in similar magnitude, with incompatibility normally being somewhat stronger. This observation mirrors the finding in the simulation of Yukilevich et al. [56], where mutations with large fitness effects caused both synergistic and antagonistic effects. The framework here introduced could easily be extended to simulate different evolution and selection scenarios to test the generality of our findings.

An experimental approach to heterosis research has been that of predicting the levels of heterosis using available genetic and phenotypic data about the parents, for example metabolic profiling [5,57]. This study showed that genetic markers, metabolic markers and more general phenotypes such as parental biomass carried complementary information necessary to predict heterosis [57]. This finding would confirm the role of epistatic effects in heterosis. Another study using a metabolic network as a representation of the individuals also concluded that epistasis plays a fundamental role in heterosis [58]. Further analysis of epistasis in our simulated hybrids could potentially shed more light on the importance of multigenic effects in determining vigour.

### Parallels between local heterozygosity effects and epistasis

The mechanisms of dominance, over-dominance and under-dominance can be considered at a single locus level (though several cases of dominance would be necessary to explain heterosis by this mechanism alone). Epistasis is the interaction between alleles of separate genes, and thus cannot be explained without at least looking at pairs of genes. However, there are notable parallels between the three mechanisms of epistasis and the local heterozygosity mechanisms. In both dominance and negative epistasis, at least one allele can be removed from the network without a loss in fitness, as the other allele of the pair compensates for its function. In both over-dominance and positive epistasis, the interaction between the two alleles leads to a greater fitness, while in the case of under-dominance or epistatic incompatibility, removing either allele increases fitness. As shown in Figures 5 and 6, the local heterozygosity effects show largely the same pattern as the epistatic effects, with the exception of over-dominance, which is strong from the start, contrary to positive epistasis.

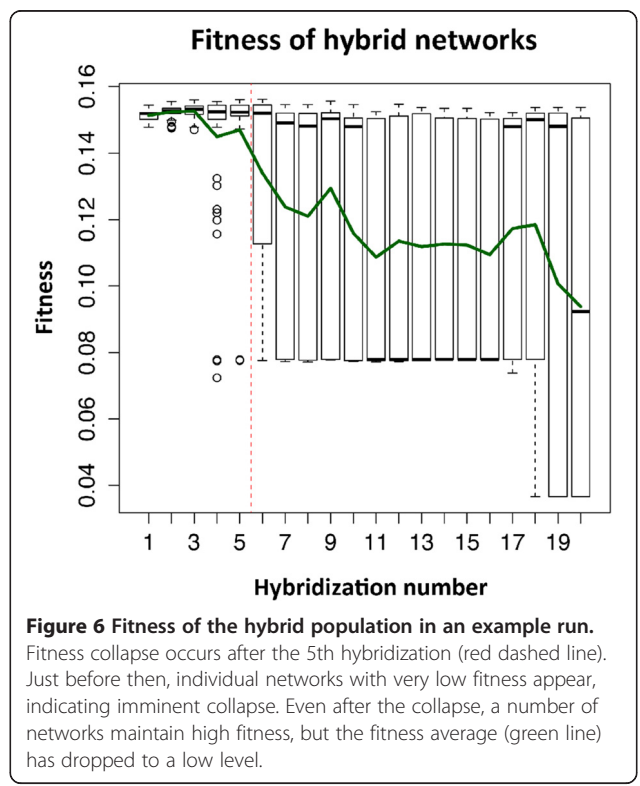

**Figure 6 Fitness of the hybrid population in an example run.** Fitness collapse occurs after the 5th hybridization (red dashed line). Just before then, individual networks with very low fitness appear, indicating imminent collapse. Even after the collapse, a number of networks maintain high fitness, but the fitness average (green line) has dropped to a low level.

## Robustness of the model to the choice of parameters

Our simulation is based on a number of parameters governing the generation of the initial networks, the choice of modules for fitness calculation, the size of each network and the size of the populations, the relative rates of different mutation mechanisms and the overall mutation rate as well as the number of generations allowed for initial adaptation. We experimented with different values for all these parameters to ensure that the results we observed were not artifacts of the specific parameters we chose. Because of the stochasticity at the center of our simulation and the computational challenge of running it under many different conditions, we have left a formal parameter space exploration as a topic for further publications.

One of the most interesting parameters is the overall mutation rate. It is very difficult to give a realistic estimate for this value in biological organisms, as it is likely to vary markedly between species (and between environments) and because only the mutations that are not removed by selection are easily observable. When the same number of mutations was spread out over a greater number of generations in the initial adjustment period, the fitness equilibrium reached by the adjusted population was higher. We observed that lower mutation rates produced more pronounced heterosis, and we interpret this to there being more time for selection to remove deleterious alleles between the appearance of new mutations events, and even weakly advantageous mutations are able to spread through the population. Also, the collapse in hybrid fitness took place within a similar number of generations, regardless of the mutation rate, suggesting that a single mutation, or a very small number of mutations, is responsible for the collapse, rather than a slow build-up of different alleles in the two populations.

We also performed simulations using two environments (rather than three) which produced substantially the same results, except for a generally higher fitness level reached by the population. For more detail, see Additional file 1: Text S3, Figures S3A and S3B.

An interesting direction for future work will be the exploration of the role of the network topology and logic functions in determining the speed of adaptation, the levels of hybrid vigor achievable and the number of generations for which hybrid vigor can be sustained. Given that our model produces hybrid vigour simply based on the network characteristics of the individuals, it could be used to test and refine models of how the topology features affect the levels of epistasis [59].

## Validation of the simulation using biological data

Although so far the model hasn't been validated with biological data, the networks that we use to represent our individuals could be easily compared to regulatory networks for different species and parallels between real organisms and our networks could be established. The topological properties of our networks are in line with the currently available biological regulatory networks and the results we presented are likely to be general enough to provide insight on the real biological systems. Reproducing *in vivo* the experiments performed in our simulation requires the breeding of separate populations of individuals in specific environments for hundreds or thousands of generations with repeated analysis of the operating genetic regulation networks, which would imply a near-impossible effort even in the simplest and fastest evolving organisms, such as bacteria. However, the analysis of the genetic makeup and phenotype in hybrids created by nature itself could be a first tool to verify how well our simulation reproduces biological hybrid vigor. For example, recent research has focused on the genetic and phenotypic characterization of large sets of wild isolates of yeast [60-63]. It would be interesting to identify within these collections sets of individuals that are the result of hybridization of strains at different points in time. This type of data could be used to establish a parallel between a real organism and our simulation, possibly leading to further biological insights.

Although excellent results have been achieved in predicting heterosis levels with a combination of genetic and metabolic information [5,57], simulations of the type presented here will be useful to test mechanistic explanations of the different aspects of hybrid vigour that underlie the observed metabolic changes. In our simulation, no specific assumption about the role of the nodes in the network is made. The general framework of the simulation can be used as a way to look for how the genetic and metabolic aspects of the organism fitness are related through a multitude of different regulatory interactions. For example, extensions to the simulation could specifically look for the role of genomic and epigenetic mechanisms that have also been implicated in heterosis [21,64].

## Conclusion

Our simulation is able to reproduce heterosis in evolving populations of gene regulatory networks. It predicts the rise in hybrid fitness as parent populations diverge, as well as the collapse in hybrid fitness due to allelic incompatibility and the instability of the heterotic fitness in subsequent crosses of hybrids. All these phenomena are observed in nature. The modelling framework allows us to measure the effects of different genetic mechanisms that may cause heterosis and shows overdominance and positive epistasis to be the strongest effects. Our model is more detailed and less abstract than previous network-based models of heterosis and lends itself to modifications and extensions that could enable the study of haploid organisms, allopolyploidy and other aspects of evolutionary theory. The emergence of genetic incompatibility between diverging population means that the simulation might even be applied to model processes of speciation.

## Methods

The simulation was written using R version 2.13.2. The following packages were used BoolNet version 1.4, igraph version 0.5 and sfsmisc version 1.0-19.

### Summary of the evolutionary algorithm

Each individual in the simulation is represented by a network composed of components inherited from two parents. The fitness of this network is measured as its response to a set of three environments, which are defined by the fixed states of certain nodes. For each environment we define a subset of nodes (a module) which is to be turned ON, while the other modules are to be turned OFF, to achieve 'perfect' fitness. Based on their fitness, networks are selected to reproduce according to a Metropolis-Hastings algorithm (see below and Additional file 1: Text S7). From each selected individual, a number of "gametes" are formed, which are lists of alleles inherited from either parent of the selected individual. Two "gametes" are combined to generate a new diploid network. For more information of each step in this cycle, see the Additional file 1: Text S4-S7, Figures S4 and S5.

### The individual: networks, Boolean logic and attractors

Our simulation uses regulatory networks in which nodes represent genes or gene products and directed edges represent regulatory interactions (Table 1). In biological terms, these interactions can result from many mechanisms such as transcription factor regulation of targets, regulation of an mRNA by an RNA binding protein, post-translational modifications in signaling cascades or any other relation that produces an effect of the product of one gene on the product of another. The initial random networks were generated to ensure their statistical resemblance to biological regulatory networks, namely they were built to display realistic degree distributions (in-degree, out-degree) and to have modules, subsets of nodes that are highly interconnected but less connected with nodes outside of the module.

**Table 1 Simulated genetics in an individual**

| Biological entities | Our simulation |
| --- | --- |
| Gene | Node |
| Gene activity (e.g. expression, post-translational modification, etc.) | Binary node state |
| Regulatory interaction | Connection (edge) between nodes |
| Allele | logic function and potential for in- and out-going connections that determine the state of a node and its function in the dynamics of the network |
| Gamete | Haploid list of alleles selected at random from a diploid individual, combined with another gamete, forms a new diploid |

The Boolean framework imposes binary values on the state of each node (gene), allowing them to be either ON or OFF. The state of a node at a further time step in the simulation is given by logic transition functions depending on the states of the nodes that the node receives inputs from. The logic functions are constructed of AND, OR and NOT relations and are allowed to mutate during evolution. These logic functions were assigned randomly to each node in the initial network.

Iteratively updating the states of the network causes the network to reach a repeated combination of states. This is the attractor of the system, which can be either a single state (fixed point) or a cyclically repeating group of states from which no more transitions are possible. Multiple attractors can exist for a single network (individual) that can be reached starting from different initial conditions (initial gene states). In our model, we used synchronous updating, i.e. all nodes were assigned their new state simultaneously, rather than one at a time. Our preliminary tests showed no major differences in the attractors generated by synchronous and asynchronous updating (see Additional file 1: Text S8), due to the fixed states of the environment nodes and the limited size of our networks. Synchronous updating is much faster, so we decided to use it for our model.

### Network generation

For the initial generation of the networks, we used an algorithm first described by Holme and Kim [65] consisting of preferential attachment and triad formation (to increase the clustering, or modularity, of the network), and adapted it for directed networks using ideas from Prettejohn et al. [66]. The resulting algorithm had two tuning parameters which allowed limited control over the network topology. These parameters were fixed such that the algorithm best reproduced observed degree distributions as well as ultra-small-world and modular properties that have been seen in biological networks [31]. Throughout evolution, the networks acquired more realistic properties, such as having a few nodes with many outgoing connections (hubs) and many more nodes with very few connections (data not shown). This suggests that evolution was not compromising the properties of the networks that were built into them in the beginning.

### Calculating phenotype and fitness

The capacity of the network to correctly respond to external conditions was used as a biologically meaningful and generic measure for fitness.

To use a simplistic example, one could define for a plant three binary parameters that can be measured for a specific environment, like day-length with regard to a threshold, infection by a pathogen and presence of frost. Environments can then be defined based on combinations of these parameters, so different environments could for example represent different seasons. The individual's phenotype emerges in response to the present environment (e.g. a tree losing its leaves in autumn).

Hence three nodes were arbitrarily selected from the initial network and defined as 'environment nodes' (see Figure 7). These nodes have fixed states, which cannot be altered dynamically during the simulation. These nodes represent factors of the environment that the organism may be able to perceive (e.g. the presence of a pathogen). Any incoming connections these nodes gained during network creation are erased but out-going connections to other nodes (which represent the organisms means of perceiving its environments, e.g. receptor proteins) are retained. The networks may evolve new connections to these environment nodes or lose old ones through subsequent mutations. A specific combination of states for these three nodes was termed an 'environment'. We perform the fitness calculation for three different environments (different combination of states of the same three environment nodes). In order to represent groups of genes involved in responding to each environment, we chose three modules from our network (sub-networks where the nodes are particularly highly interconnected within the sub-network) and assigned each to one environment. We then specified that under a specific environment, the ideal network response (i.e. the ideal phenotype) is achieved when in the network attractor all the nodes of the corresponding module are turned ON, whilst all the other nodes are OFF.

To calculate the fitness under one environment, a single value for each module is formed by averaging the states of the nodes contained in the module. Then the average of the values of all the modules that are not associated with this specific environment is subtracted from the value of the associated environment. This is repeated for each environment.

The average fitness in all three environments is the overall fitness of the individual. This resulted in a fitness value between 1 (a perfectly responding network), and −1 (a network that responds in exactly the wrong way to every environment). We chose this method of defining fitness rather than the related, but somewhat simpler methods used in [67] and [19], because it allows greater freedom in the network evolution, including the complete loss and duplication of nodes. For more information, see the Additional file 1: Text S2.

### Selection based on fitness

Selection was then performed to choose the individuals that will produce offspring in the following generation (Table 2). The probability of individuals being selected increased with their fitness, such that the population evolved better adaptation to the environments. For this, we used a Metropolis-Hastings algorithm (see Additional file 1: Text S7). This could be visualized as climbing to the top of mountains in the evolutionary fitness landscape, through small steps that increase fitness. However, to prevent the population from getting trapped onto local peaks from which the fitness cannot increase any further, low fitness individuals were also included in the selection, with much lower probability, allowing the population to descend towards a valley that might allow it to climb up a different, potentially higher, mountain.

To go back to our example, plants that can adapt their phenotype to the seasons correctly will stand a greater chance of surviving and generating offspring, which will lead to better and better adaptation of the population to its environment.

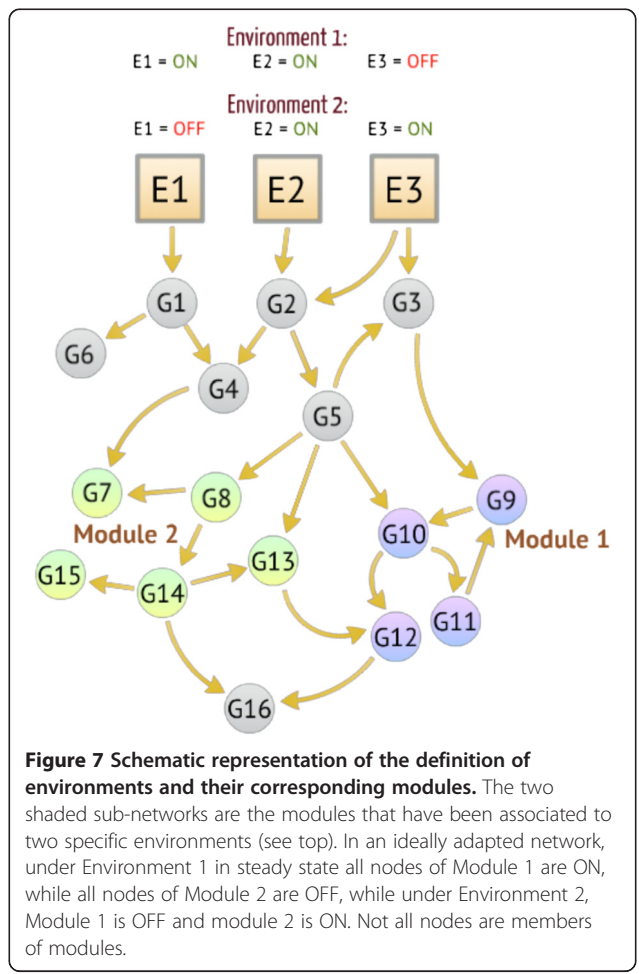

**Figure 7 Schematic representation of the definition of environments and their corresponding modules.** The two shaded sub-networks are the modules that have been associated to two specific environments (see top). In an ideally adapted network, under Environment 1 in steady state all nodes of Module 1 are ON, while all nodes of Module 2 are OFF, while under Environment 2, Module 1 is OFF and module 2 is ON. Not all nodes are members of modules.

**Table 2 Population simulation**

| Population | Group of 40 networks (reduced to 10 in every round of selection) |
|---|---|
| Regulatory dynamics | The network passes through different states depending on environment and wiring, until it arrives at a stable or recurring attractor state |
| Phenotype | Steady state or state cycle of the states of the network's nodes (attractor) caused by the network dynamics under a certain environment |
| Fitness | Similarity between phenotype and ideal phenotype in a specific environment |

Initially, the definition which module should be switched ON in each environment is entirely arbitrary. This only serves to provide a highly imperfect initial situation that allows adaptation by the evolutionary algorithm and is not intended to represent a situation found in biological evolution. At the end of the initial adaptation phase, however, the networks have evolved to process the inputs (i.e. states of the externally controlled environment nodes) into environment-specific outputs. Each module can now be seen as representing a fitness-enhancing (i.e. biologically useful) response to a specific environment.

### Calculation of local heterozygosity effects

The initial networks were generated with 22 nodes plus three nodes with only outgoing connections, which represent environmental factors (environment nodes). Over the course of evolution, genes can be lost or duplicated, so the number of nodes in each diploid network varies. In addition, each heterozygotic locus is represented by two nodes, while homozygotic loci are only represented by one node. This is a valid simplification, as two identical nodes have the same effect on network dynamics as a single node.

If a network included two identical alleles, they would have the same inputs and the same logical function. As a consequence they would always be assigned the same state, both ON or both OFF. In addition, they would have the same outputs, so any other node that receives an input from them will receive inputs from both, connected by an OR-gate. However, as the state of the two nodes is always the same, the result of the OR gate will always be the same as the state of the two nodes. Thus, by representing homozygous loci by only one node, the computation can be simplified without affecting the network dynamics.

We define an allele as the information describing one node in a network. This includes the genes it can receive inputs from and the genes it can output to and the logical function used to assign the node's state.

To examine dominance, we considered each heterozygous gene in a network, i.e. each gene for which the alleles inherited from each parent were not identical. For each gene, we measured network fitness in the complete network as well as the two networks in which the alleles from either parent were removed, as described above, these networks represent individuals that are homozygotic for the other allele, respectively. We defined a dominance effect as the difference between the fitness of the complete (heterozygotic) network and the average fitness of the two homozygotes. The network dominance score is then the sum of scores of all heterozygous loci at which dominance occurs. To exclude numerical error, only effects of more than 0.5% of the whole network fitness were counted. Only cases where the complete network fitness was greater than the average of the homozygotes were counted towards the dominance effect. If the fitness of the complete network was lower than the average of the homozygotes (e.g. if the dominant allele was deleterious) the effect was counted as under-dominance (see below).

Over-dominance was measured as the difference between the complete network and the average fitness of the homozygotes, in cases where the fitness of both homozygotic networks was lower than the fitness of the complete network.

Under-dominance describes the opposite of over-dominance, i.e. a case in which the presence of two different alleles of one gene causes lower fitness than the presence of either allele in homozygous form. Thus, under-dominance is opposed to heterosis and is one of the mechanisms involved in the collapse of hybrid fitness and subsequent hybrid incompatibility. For examples of these mechanisms, see the Additional file 1.

### Calculating the effects of epistasis

To measure epistasis, we need to assess the fitness effect of an allele in different genetic backgrounds, i.e. in the presence or absence of another allele of a different locus.

Consider a diploid network that contains the alleles G1, G1', G2 and G2', where G1 and G2 have been inherited from one parent and G1' and G2' from the other. In order to measure epistasis between the alleles G1 and G2'*, we estimate the fitness the network would be expected to have if the two alleles were independent (i.e. no epistasis) and compare it to the fitness of the actual network with all alleles present. We calculate the fitness of the complete network, the network from which G1 was removed (i.e. where G1' is homozygous, but both G2 and G2' are present), the network where G2' is removed (homozygous for G2, but both G1 and G1' are present) and the network with neither G1 nor G2' (homozygous for G1' and G2).

The epistasis between the two alleles G1 and G2' is calculated as the difference between the fitness of the complete (double heterozygote) network and the predicted network fitness in the absence of epistasis: the product of the fitness values of the two single homozygote networks divided by the fitness of the double homozygote network.

$$ep_{G1xG2'} = F_{complete} - \frac{F_{w/o\ G1} * F_{w/o\ G2'}}{F_{w/o\ G1,G2'}}$$

*These calculations are always done for alleles of different loci. G1 and G1' are an allelomorphic pair, so their interaction would fall under dominance, overdominance or underdominance. G1 and G2 are inherited from the same parent. While there may be epistasis between them, it would also be present in the parent that the two alleles are from, so it is irrelevant for heterosis. The same applies to epistasis where at least one of the alleles

involved is homozygous. The interaction between G1' and G2 has to be calculated separately.

An epistatic interaction was classed as positive, if the fitness of the double heterozygote network was greater than the predicted one. One mechanism would be that the two loci were involved in some common function and the absence of a particular allele at one locus disrupts that function. As such positive epistatic interactions increase the fitness of the hybrid, they can generate heterosis.

If the measured fitness of the double mutant is lower than the prediction, either negative epistasis (as defined by [68]) or epistatic incompatibility (c.f. [69,70]) is present. In negative epistasis the two alleles have redundant functions, so removing either one does not affect fitness but removing both does. Negative epistasis cannot cause heterosis, since both parents would have the function from the allele they handed down to the hybrid. But it can prevent heterosis from occurring, e.g. if the interaction exists between two dominant alleles.

Alternatively, having both alleles could disrupt some function. In this case, the fitness of both single mutants is higher than the fitness of the complete network, leading to an even higher prediction for the double mutant. This would be a case of epistatic incompatibility, which is strongly opposed to heterosis. This kind of interaction appears to be the prime cause of the fitness collapse observed in the hybrids.

## Availability of supporting data

We make the code available to the community through a GitHub repository at https://github.com/VeraPancaldi/Heterosis-GRN-simulation and hope that it will become a useful tool for further computational analysis in the field. This version may be improved or expanded in the future. In addition, we provide a static version of the source code on figshare under http://figshare.com/articles/Heterosis_GRN_simulation/1247499.

## Additional file

**Additional file 1: Text S1.** Examples of heterosis by different mechanisms. **Text S2.** Calculating network fitness. **Text S3.** Simulation results using two environments. **Text S4.** Generating alleles and building diploid networks. **Text S5.** Implementation of mutations. **Text S6.** Implementation of independent assortment. **Text S7.** Implementation of selection. **Text S8.** Synchronous versus asynchronous updating.

## Competing interests
The authors declare that they have no competing interests.

## Authors' contributions
PMFE and HER designed the algorithm and wrote the code for the core simulation. HER performed the analysis of network structures. PMFE performed the simulation runs, and participated in drafting the manuscript and analysed the resulting data. VP conceived this study and drafted the manuscript. All authors have approved the final manuscript.

## Acknowledgements
This work was supported by the European Commission FP7 Revolution project (grant number 233325) and a FEBS fellowship to V.P. and was partially carried out by P. M. F. E. and H. E. R. as a project for the Part III course in Systems Biology at Cambridge University supervised by V.P. in the laboratory of David C. Baulcombe. We thank David C. Baulcombe and Krystyna A. Kelly for critical reading of the manuscript.

## Author details
[1]Department of Plant Sciences, University of Cambridge, CB2 3EA Cambridge, UK. [2]Current address: John Innes Centre, Norwich Research Park, Norwich NR4 7UH, UK. [3]Current address: The Nuffield Department of Clinical Medicine, Oxford University, Peter Medawar Building for Pathogen Research, Oxford OX1 3SY, UK. [4]Current address: Structural Biology and BioComputing Programme, Spanish National Cancer Research Centre (CNIO), Calle Melchor Fernández Almagro, 3, Madrid E-28029, Spain.